\documentstyle[12pt,epsf]{article}
\newcommand{\be}{\begin{equation}}
\newcommand{\ee}{\end{equation}}
\newcommand{\bea}{\begin{eqnarray}}
\newcommand{\eea}{\end{eqnarray}}
\newcommand{\bt}{\begin{tabular}}
\newcommand{\et}{\end{tabular}}
\newcommand{\ba}{\begin{array}}
\newcommand{\ea}{\end{array}}

\newcommand{\gapproxeq}{\lower .7ex\hbox{$\;\stackrel{\textstyle >}{\sim}\;$}}
\newcommand{\lapproxeq}{\lower .7ex\hbox{$\;\stackrel{\textstyle <}{\sim}\;$}}
\def\ne{\hbox{$\nu_e$ }}
\def\nm{\hbox{$\nu_\mu$ }}
\def\nt{\hbox{$\nu_\tau$ }}
\def\ane{\hbox{$\overline{\nu}_e$ }}
\def\anm{\hbox{$\overline{\nu}_\mu$ }}
\def\ant{\hbox{$\overline{\nu}_\tau$ }}
\def\neL{\hbox{$\nu_{e L}$ }}
\def\nmL{\hbox{$\nu_{\mu L}$ }}
\def\ntL{\hbox{$\nu_{\tau L}$ }}
\def\aneR{\hbox{$\overline{\nu}_{e R}$ }}
\def\anmR{\hbox{$\overline{\nu}_{\mu R}$ }}
\def\antR{\hbox{$\overline{\nu}_{\tau R}$ }}
\def\dm2{\hbox{$\Delta m^2$ }}

\def\s2t{\hbox{$\sin (2 \theta$) }}
\def\ss2t{\hbox{$\sin^2 (2 \theta)$ }}
\def\c2t{\hbox{$\cos (2 \theta)$ }}
\def\cc2t{\hbox{$\cos^2 (2 \theta)$ }}

\def\emdm2{\hbox{$\Delta m^2_{e \mu}$ }}

\def\ems2t{\hbox{$\sin (2 \theta_{e \mu})$ }}
\def\emss2t{\hbox{$\sin^2 (2 \theta_{e \mu})$ }}
\def\emc2t{\hbox{$\cos (2 \theta_{e \mu})$ }}
\def\emcc2t{\hbox{$\cos^2(2 \theta_{e \mu})$ }}

\def\etdm2{\hbox{$\Delta m^2_{e \tau}$ }}

\def\ets2t{\hbox{$\sin(2 \theta_{e \tau})$ }}
\def\etss2t{\hbox{$\sin^2 (2 \theta_{e \tau})$ }}
\def\etc2t{\hbox{$\cos(2 \theta_{e \tau})$ }}
\def\etcc2t{\hbox{$\cos^2 (2 \theta_{e \tau})$ }}

\def\ene{\hbox{$\nu_e \rightarrow \nu_{e}$ }}
\def\aenae{\hbox{$\overline{\nu}_e \rightarrow \overline{\nu}_{e}$ }}
\def\enae{\hbox{$\nu_e \rightarrow \overline{\nu}_{e}$ }}
\def\aene{\hbox{$\overline{\nu}_e \rightarrow \nu_{e}$ }}

\def\noe{\hbox{$n^0_{\nu_e}$ }}
\def\nom{\hbox{$n^0_{\nu_{\mu}}$ }}
\def\anot{\hbox{$n^0_{\nu_{\tau}}$ }}
\def\nox{\hbox{$n^0_{\nu_x}$ }}
\def\anoe{\hbox{$n^0_{\overline{\nu}_e}$ }}
\def\nne{\hbox{$n_{\nu_e}$ }}
\def\nnm{\hbox{$n_{\nu_{\mu}}$ }}
\def\nnt{\hbox{$n_{\nu_{\tau}}$ }}

\def\anne{\hbox{$n_{\overline{\nu}_e}$ }}
\def\annm{\hbox{$n_{\overline{\nu}_{\mu}}$ }}
\def\annt{\hbox{$n_{\overline{\nu}_{\tau}}$ }}

\begin{document}
\setcounter{page}{1}
\thispagestyle{empty}
\baselineskip=20pt
\hfill{
\begin{tabular}{l}
DSF$-$17/97 \\
hep-ph/9704374
\end{tabular}}

\bigskip

\begin{center}
{\Large \bf Three Flavour Majorana Neutrinos with Magnetic Moments
in a Supernova}
\end{center}

\vspace{1cm}

\begin{center}
{\large S. Esposito, V. Fiorentino, G. Mangano and G. Miele} 
\end{center}

\vspace{1cm}

\normalsize
\noindent
{\it Dipartimento di Scienze Fisiche, Universit\`a di Napoli 
''Federico II'', and INFN, Sezione di Napoli, Mostra D'Oltremare Pad. 20, 
I-80125 Napoli, Italy.}

\vspace{1cm}

\begin{abstract}
The resonant transition effects MSW and NSFP for three flavour
Majorana neutrinos in a supernova are considered, where the transition
magnetic moments are likely to play a relevant role in neutrino
physics. In this scenario, the deformed thermal neutrino distributions
are obtained for different choices of the electron-tau mixing angle.
Detailed predictions for the future large neutrino detectors are also
given in terms of the ratio between the spectra of recoil electrons
for deformed and undeformed spectra. 
\end{abstract}

\vspace{1truecm}

\normalsize
\newpage

\section{Introduction}
\indent
Though the Standard Model (SM) for fundamental interactions has
received a striking confirmation from experiments, the neutrino sector
still remains a puzzling subject for theoretical speculations and for
conceiving new experiments. The determination of the possible neutrino
masses and the consequent mixing angles is the main goal of this huge
scientific effort. 

Several accelerator experiments have analyzed a quite wide range of
the space of neutrino parameters by putting a certain number of upper
bounds on these physical quantities \cite{2}. However, it seems that
the most interesting and promising observations should come from the
cosmological and astrophysical side, which enables to detect extremely
weak phenomena as neutrino oscillations. 

In this concern, the new experimental results on solar neutrinos seem
to confirm the previously observed deficit of their flux \cite{3}. On
this ground, the so--called solar neutrino problem can be reasonably
considered as a well--established experimental observation. This, of
course, has a number of implications on the previous question,
concerning the fundamental parameters of neutrino physics (masses,
mixing angles, e.m. properties). In particular, the solution of
solar neutrino problem needs massive neutrinos and not vanishing
mixing angles. 

The mechanisms proposed to explain the deficit in the observed number
of solar neutrinos with respect to the theoretical predictions of the
Standard Solar Model (SSM) \cite{4} are basically three. As far as the
pure oscillation in vacuum is concerned, it is strongly disfavoured by
the experimental evidence for a quite strong energy dependence of the
flux depletion mechanism \cite{5}. Certainly, the most promising
solution of the solar neutrino problem is provided by the resonant
oscillation mechanism, the Mikheyev-Smirnov-Wolfenstein effect (MSW)
\cite{6}, which has the correct energy dependence. Nevertheless, one
can imagine the superposition of a Neutrino Spin-Flavour Precession
(NSFP) mechanism too \cite{7}. This effect is relevant when a strong
transverse magnetic field is present and if the magnetic dipole
moments of neutrinos are large enough. 

The nature of neutrinos as Dirac or Majorana particles is also still
unknown. In scenarios proposed by GUT theories as $SO(10)$ \cite{8},
neutrinos are described by Majorana spinors. In this framework, since
new physics is present beyond SM, the e.m. properties can also get a
strong enhancement. This is in particular the case of some extended
unified theories \cite{11}. 

In this paper we extend the analysis recently performed by some of the
present authors \cite{10}. We will consider three-flavour Majorana
neutrinos for which, due to CPT invariance, only flavour-changing
magnetic dipole moments are, in general, non vanishing \cite{11}. The
resonant neutrino oscillations are considered in a supernova where the
conditions of large density and strong magnetic field are likely to be
obtained. 

The paper is organized as follows. In section 2 we consider the
effective Hamiltonian describing the interactions of three flavour
Majorana neutrinos in matter. In particular we study the MSW and NSFP
effects as a function of the mass difference and mixing angles. In
section 3, the main supernova characteristics in terms of radial mass
density profile and magnetic field are reviewed, and the adiabaticity
conditions are studied. The outcoming thermal neutrino spectra are
then obtained as functions of survival and transition probabilities.
In section 4 the predictions about deformed thermal neutrino spectra
are obtained, and the ratio between the spectra of recoil electrons
for the deformed neutrino distributions and for the unmodified ones is
computed. Finally, in section 5 we give our conclusions. 

\section{Resonant transitions of Majorana neutrinos}

For three flavour Majorana neutrinos the independent degrees of
freedom, which do not take mass at the intermediate scale
($10^{11}~GeV$) \cite{8} and are coupled with weak interaction, can be
denoted by $\neL$, $\nmL$, $\ntL$ and for antineutrinos by $\aneR$,
$\anmR$, $\antR$. For simplicity, hereafter we will omit the
indication of chirality being clear that it is left-handed for
neutrinos and right-handed for antineutrinos. 

In presence of a transverse magnetic dipole moment a neutrino can flip
its spin and thus change its chirality. However, CPT invariance
forbids these transitions for Majorana neutrinos, unless they change
flavour at the same time, namely $\nu_{\alpha L} \rightarrow
\overline{\nu}_{\beta R}$, with $\alpha,\beta=e,\mu,\tau$ and $\alpha
\neq \beta$.\\ This kind of transitions are called Neutrino
Spin--Flavour Precessions (NSFP) \cite{nsfp}, and their probability
can be large if the amplitudes of the transverse magnetic field and of
the off-diagonal dipole magnetic moment are large enough. 

As neutrino oscillations in matter (MSW), even the NSFP can receive a
resonant enhancement from the presence of a dense medium \cite{nsfp}.
This is for example the case of stellar matter or the extreme
condition of a supernova. 

Let us denote with $\nu$ the vector in the flavour space $\nu \equiv
(\ne, \nm, \nt)$, and analogously $\overline{\nu} \equiv (\ane, \anm,
\ant)$. The $3\times 3$ unitary mixing matrix can be written as
\cite{12} 
\be
U = \left(\begin{array}{ccc} C_\varphi C _\omega & C_\varphi S_\omega &
S_\varphi \\ - C_\psi S_\omega e^{i \delta} - S_\psi S_\varphi C_\omega
e^{-i \delta} & C_\psi C_\omega e^{i \delta} - S_\psi S_\varphi S_\omega
e^{-i \delta} & S_\psi C_\varphi e^{- i \delta} \\
S_\psi S_\omega e^{i \delta} - C_\psi S_\varphi C_\omega
e^{-i \delta} & - S_\psi C_\omega e^{i \delta} - C_\psi S_\varphi S_\omega
e^{-i \delta} & C_{\psi} C_{\varphi} e^{-i \delta} \end{array} \right)~~~.
\label{2.1}
\ee
We will assume for simplicity $\delta=0$, and thus $U$ represents the
mixing matrix for antineutrinos as well. 

The evolution equation for neutrino wave functions travelling along
the radial coordinate $r \simeq c t$ is 
\be
i {d \over dr} \left( \begin{array}{c} \nu \\ \overline{\nu}\end{array}
\right) = H \left( \begin{array}{c} \nu \\ \overline{\nu}\end{array}
\right) = \left(\begin{array}{cc} H_0 & B_{\perp}M \\
-B_{\perp}M & \overline{H}_{0} \end{array} \right)
\left( \begin{array}{c} \nu \\ \overline{\nu}\end{array}
\right)~~~.
\label{2.2}
\ee
The symmetric matrix $H_0$ is the $3 \times 3$ hermitian effective
Hamiltonian ruling the resonant flavour transition in the flavour
basis; it  is given by 
\begin{equation}
H_0 ={ 1 \over 2E} U \left(\begin{array}{ccc}
m_1^2 & 0  &  0\\ 0 & m_2^2 & 0 \\ 0 & 0 & m_3^2 \end{array}\right)
U^{\dag} + 
\left(\begin{array}{ccc}
N_1(r) & 0  &  0\\ 0 & N_2(r) & 0 \\ 0 & 0 & N_2(r) 
\end{array}\right)~~~,
\label{2.3}
\end{equation}
where $N_1(r) = \sqrt{2} G_F (N_e - N_n/2)$, $N_2(r)= - G_F 
N_n/\sqrt{2}$, $N_e$ and $N_n$ being the electron and neutron number
density, respectively. Substituting (\ref{2.1}) in (\ref{2.3}) we get
\begin{eqnarray}
\left(H_0\right)_{ee} & = & \Sigma + N_1(r)
-\Lambda S^2_{\varphi} - \Delta C_{2 \omega}C_{\varphi}^2~~~,
\nonumber\\
\left(H_0\right)_{e\mu} & = & 
- {\Lambda \over 2} S_{2 \varphi} S_{\psi}  
+\Delta\left(S_{2 \omega} C_{\varphi} C_{\psi} + C_{2 \omega} S_{\varphi}
S_{\psi}\right)~~~,
\nonumber\\
\left(H_0\right)_{e\tau} & = & 
- {\Lambda \over 2} S_{2 \psi} C_{\varphi}  
-\Delta\left(S_{2 \omega} C_{\varphi} S_{\psi} -{1 \over 2} 
C_{2 \omega} S_{2 \varphi} C_{\psi}\right) ~~~,
\nonumber\\
\left(H_0\right)_{\mu \mu} & = &  \Sigma + N_2(r)
- \Lambda C^2_{\varphi} S^2_{\psi} - \Delta
\left[S_{2 \omega} S_{2 \psi} S_{\varphi} + C_{2 \omega} 
\left(S^2_{\varphi} S_{\psi}^2 - C_{\psi}^2 \right) \right] ~~~,
\nonumber\\
\left(H_0\right)_{\mu \tau} & = & 
- {\Lambda \over 2} S_{2 \psi} C^2_{\varphi} - \Delta
\left[S_{2 \omega} C_{2 \psi} S_{\varphi} + {1 \over 2} C_{2 \omega} 
S_{2 \psi}\left(1+S^2_{\varphi}\right) \right] ~~~,
\nonumber\\
\left(H_0\right)_{\tau \tau} & = & \Sigma + N_2(r)
- \Lambda C^2_{\psi} C^2_{\varphi} + \Delta
\left[S_{2 \omega} S_{2 \psi} S_{\varphi} + C_{2 \omega}
\left(S_{\psi}^2 - C_{\psi}^2 S_{\varphi}^2\right) \right] 
~~~,
\label{2.4}
\end{eqnarray}
with $\Sigma = (m_1^2 + m_2^2)/4E$, $\Delta=(m_2^2 - m_1^2)/4E$ and
$\Lambda=(m_1^2 + m_2^2 - 2 m_3^2)/4E$. The corresponding Hamiltonian
for antineutrinos, denoted with $\overline{H}_0$, is obtained from
(\ref{2.4}) by replacing $N_{1}(r), ~N_{2}(r) \rightarrow -N_{1}(r),
~-N_{2}(r)$. In what follows the diagonal contribution in (\ref{2.4})
proportional to $\Sigma$ can be neglected since it only gives a common
phase factor. 

The quantity $M$ is the matrix of magnetic dipole moments
\be
M = \left(\begin{array}{ccc} 0  & \mu_{e\mu} & \mu_{e \tau}\\
-\mu_{e \mu} & 0 & \mu_{\mu \tau} \\
-\mu_{e \tau} & -\mu_{\mu \tau} & 0 \end{array} \right)~~~.
\label{2.5}
\ee
The resonant conditions for transformations $\ne \leftrightarrow \nm,
~\nt$ and for $\ne \leftrightarrow \anm,~\ant$  can be obtained by
requiring the approaching of two different eigenvalues of $H$ in
Eq.(\ref{2.2}). For small mixing angles this essentially coincides
with the condition of having coincident diagonal elements, namely,
neglecting second order terms in $S_\omega$,$S_\psi$ and $S_\phi$

\begin{eqnarray}
&N_1(r_s)-N_2(r_s) = 2 \Delta ~~~~~~~~~~~~~~~~~~&(\ne \leftrightarrow \nm)~~~,
\label{2.6}\\
&N_1(r_s)-N_2(r_s) = 
\Delta-\Lambda ~~~~~~~~~~~~~~~~~~&(\ne \leftrightarrow \nt)~~~,
\label{2.7}\\
&N_1(r_s)+N_2(r_s) = 2 \Delta ~~~~~~~~~~~~~~~~~~&(\ne \leftrightarrow \anm)~~~,
\label{2.8}\\
&N_1(r_s)+N_2(r_s)=
\Delta-\Lambda ~~~~~~~~~~~~~~~~~~&(\ne \leftrightarrow \ant)~~~.
\label{2.9}
\end{eqnarray}
In the same way, the resonant conditions for $\ane \leftrightarrow
\anm, ~\ant$ and for $\ane \leftrightarrow \nm,~\nt$ are obtained from
the charge conjugate transitions (\ref{2.6})-(\ref{2.9}) by changing
sign in the corresponding right hand sides. Note that $r_s$ denotes
the value of coordinate $r$ for which each resonant condition is
verified. 

Once that neutrino parameters, namely mixing angles $\psi$, $\varphi$
and $\omega$, and $\Delta$ and $\Lambda$ are fixed, one can study the
occurrence of resonant conditions (\ref{2.6})--(\ref{2.9}) and charged
conjugate ones, by using the particular density profile for electrons
and neutrons, contained in $N_1(r)$ and $N_2(r)$. 

In Ref.\cite{13} the above description for neutrino dynamics in
presence of matter has been used in order to explain the solar
neutrino problem. The authors assume there for simplicity that flavour
mixing occurs in the $e$--$\mu$ sector only. This corresponds to take
in Eq.(\ref{2.1}) $\psi,\varphi<<\omega$ and $\delta=0$. Thus, since
$\nt$ and $\ant$ are completely decoupled, one can consider in
(\ref{2.3}) only the $2 \times 2$ matrix involving $e$ and $\mu$
labels. 

To be predictive about neutrino fluxes, one would need the precise
expression for the solar magnetic field. Unfortunately, since 
this form is quite unknown till now, the authors consider several
magnetic field configurations treated in literature. 

As a result of their analysis, it is shown that for a small value of
the mixing angle $\theta_{e \mu}=\omega$, $\ne \leftrightarrow \anm$
transitions are sufficient to account for the solar neutrino problem.
In this case the predictions strongly depend on the magnetic field
configurations even if a best fit is achieved for the so-called LIN2
\cite{13} solar magnetic field parametrization. 
Typical values of neutrino parameters able to reproduce the data are
$\Delta m^2_{e \mu} \simeq \left(10^{-8} \div 10^{-7} \right)~eV^2$
and $\sin\left(2 \theta_{e \mu} \right)\lapproxeq 0.2 \div 0.3$
\cite{13}. 

\section{Thermal neutrino spectra in a supernova}

The neutrino flux dynamics contained in Eq.s(\ref{2.2})-(\ref{2.4})
can be also applied to the case of a supernova. In a previous paper
some of the present authors \cite{10} already considered this situation,
but in the case of negligible magnetic moments. In the following
we extend that analysis in order to consider the flavour changing
magnetic moment terms too.

In a supernova the mass density ranges from $\sim 10^{-5} g/cm^3$ in
the external envelope up to $\sim 10^{15} g/cm^3$ in the dense core.
We will assume, hereafter, the following density profile \cite{14} 
\be
\rho \; \simeq \; \rho_0 \, \left( \frac{R_0}{r} \right) ^3~~~,
\label{3.1}
\ee
with $\rho_0 \simeq 3.5 \times 10^{10} g/cm^3$, and $R_0 \simeq 1.02
\times 10^7 cm$. The quantity $R_{0}$ denotes the radius of the
so-called {\it neutrinosphere}, which represents the bounding surface
of the region in which neutrinos of a given flavour are in thermal
equilibrium. Note that, the electron fraction number $Y_e$ outside the
neutrinosphere can be assumed almost constant and fixed at the value 
$Y_e = 0.42$. 

As far as the transverse magnetic field is concerned, one can safely
assume the simple expression \cite{15} 
\begin{equation}
B_{\perp}(r) \simeq B_0 \left({R_B \over r}\right)^2~~~,
\label{3.2}
\end{equation}
where $r$ denotes the radial coordinate, and the constant $B_0 R_B^2 
\simeq 10^{24} \mbox{Gauss cm}^2$. 

Here, we will assume that inside the neutrinosphere the resonance
conditions are not satisfied, as suggested by predictions on neutrino
mass spectrum of a wide range of unified gauge models. In first
approximation neutrinos are therefore emitted from this surface with a
Fermi-Dirac distribution 
\be
n_{\nu_\alpha}^0(E) \; \simeq \; \frac{0.5546}{T_\alpha^3} \, E^2 \left[
1+\exp\left(\frac{E}{T_\alpha}\right) \right]^{-1}~~~,
\label{3.2bis}
\ee
with the different flavours equally populated. The index $\alpha$ in
Eq.(\ref{3.2bis}) denotes the particular neutrino species. Since the
production and scattering cross-sections for electron-neutrinos are
larger than for the other flavours, \ne are produced a bit more
copiously with respect to the other ones. Thus, their neutrinosphere
is larger than that for \nm, \nt. This implies that the temperature
$T_\alpha$ in (\ref{3.2bis}) for the \ne-sphere is lower than the one
of \nm, \nt. Furthermore, since \nm and \nt are produced and scatter
on the surrounding matter only through neutral currents, they have
identical spectra. Obviously, since \ne, \nm, \nt and \ane, \anm, \ant
are produced in pairs, the magnitude of neutrino and antineutrino
distributions are equal for each flavour. For the temperature of \ne
and \nm, \nt neutrinosphere we adopt the typical values $T_{e} \simeq
3~MeV$ and $T_{\mu}=T_{\tau} \simeq 6 ~MeV$. As far as the \ane
distribution is concerned, it is characterized by $T_{\bar{e}} \simeq
4~MeV$, whereas for \anm and \ant we have $T_{\bar{\mu}}
=T_{\bar{\tau}} \simeq 6~MeV$. 

Assuming that the solar neutrino problem is solved in terms of
$\ne \rightarrow \anm$ NSFP, as explained in Ref.\cite{13}, from 
Eq.s(\ref{2.6})--(\ref{2.9}) we see that in a supernova
(for $Y_e = 0.42$ outside the neutrinosphere) four resonance
conditions can be fulfilled.
With decreasing density, and for $\tau$ neutrinos more massive than
$\mu$ ones, we first encounter the region for $e$--$\tau$ resonance
transitions, and then that for $e$--$\mu$ ones. In order we have:
$\ane \leftrightarrow \nt$, $\ne \leftrightarrow \nt$, $\ane
\leftrightarrow \nm$ and $\ne \leftrightarrow \nm$. More in detail, in
the following diagram, the MSW and NSFP transitions occurring in a
supernova are reported. 

\be
\ne \, \left\{
\ba{l}
-------------------------~->  \ne\\   
---------> \nt --------------->  \nt \\
~-------------------~> \nm ----> \nm 
\ea \right. 
\nonumber
\label{3.3}
\ee
\be
\nm \, \left\{
\ba{l}
-------------------------~-> \nm\\ 
---------------> \ane ---------> \ane\\  
-------------------->  \ne ----> \ne  
\ea \right.
\nonumber
\label{3.4}
\ee
\be
\nt \, \left\{
\ba{l}
-------------------------~-> \nt\\
----> \; \ane --------------------> \ane \\              
~~~~~~~~~~~~~~~~~~~~---------> \nm ---------> \nm \\ 
~~~~~~~~~~~~~~~~~~~~~~~~~~~~~~~~~~~~~~~~~~~~~~~~~~~~~~~~
-->  \ne  ---->  \ne \\   
---------> \ne ---------------> \ne \\  
~~~~~~~~~~~~~~~~~~~~~~~~~~~~~~~~~~~~--------> \nm ----> \nm 
\ea \right.
\nonumber
\label{3.5}
\ee
\be
\ane \, \left\{
\ba{l}
-------------------------~-> \; \ane\\ 
---------------> \; \nm \; --------> \; \nm\\
~~~~~~~~~~~~~~~~~~~~~~~~~~~~~~~~~~~~~~~~~~~~~~~~~~~~~~~~
-->  \ne  ---->  \ne \\   
----> \; \nt --------------------> \nt \\        
~~~~~~~~~~~~~~~~~~~~-~-> \ne ---------> \nm  ----> \nm \\   
~~~~~~~~~~~~~~~~~~~~~~~~~~~~~~~~~~---------------~> \ne  
\ea \right.
\nonumber
\label{3.6}
\ee

To deduce the total transition probabilities, it is important to
establish if the different resonance regions overlap, namely if the
resonance widths are larger than their separation in $r$. In many GUT
models it is natural to expect $m_{\nm} << m_{\nt}$. In this paper we
make this assumption, so that the resonance regions involving
$e$--$\tau$ flavours and those involving $e$--$\mu$ flavours are well
separated between them. For example, for $10~MeV$ neutrinos with
$m_{\nm} \simeq 10^{-3}~eV$ and $m_{\nt}\simeq 10~eV$ we have the
$e$--$\tau$ resonances around the region of density $\approx
10^8~g/cm^3$ (deep in the supernova); on the contrary, the resonances
$e$--$\mu$ is in the external envelope of supernova ($\approx
1~g/cm^3$). Further, 
the MSW-NSFP resonance regions non-overlapping
condition is given by \cite{7} 
\begin{equation}
L_\rho \left| \tan{2 \theta_{\alpha \beta}} \right| 
+ \left| { 2 \mu_{\alpha \beta} B_{\perp}(r_1) \over (\Delta
m_{\alpha \beta}^2 / 2 E)} \right| L_\rho \lapproxeq r_2 - r_1~~~,
\label{3.7}
\end{equation}
where $r_1$ is the radial position of the NSFP resonance, while $r_2$
is that of the MSW one, and $L_\rho \equiv - \left[(1/\rho) d \rho/dr
\right]^{-1}$ is assumed to vary slowly between $r_1$ and $r_2$. In
Eq.(\ref{3.7}) $\alpha$ and $\beta$ label the flavours involved in the
resonance ($\alpha, \beta=e,\mu,\tau$), $\theta_{\alpha \beta}$ denote
the mixing angles, and  $\Delta m^2_{\alpha \beta}$ is the squared
mass differences of the relevant mass eigenstates. For supernova
neutrinos ($E \approx 0 \div 50~MeV$), assuming (\ref{3.1}) and
(\ref{3.2}), it is easy to see that for all models considered below,
the non-overlapping condition (\ref{3.7}) is well satisfied for both
$\ne,\ane \leftrightarrow \nt$ and $\ne,\ane \leftrightarrow \nm$
transitions. Thus, we can conclude that the four resonances occurring
in a supernova are quite well separated one each other, and as we shall see,
this implies a considerable simplification, since each resonance can
be considered independently from the others. 

For the MSW resonance we
have a simple semi--analytical formula for the survival probability
\cite{12,17} 
\begin{equation}
 P_{\alpha \beta} \left(\nu_{\alpha} \rightarrow \nu_{\alpha}\right) =
{1 \over 2} + \left( \frac 12 - 
\exp\left\{- { \pi \over 2} \gamma_{\alpha \beta} F_{\alpha \beta} \right\}  
\right) \cos{2 \theta_{\alpha \beta}} 
\cos{2 \theta_{\alpha \beta}^m}~~~,\label{3.8}
\end{equation}
where the adiabaticity parameter $\gamma_{\alpha \beta}$ is given by  
\begin{equation}
\gamma_{\alpha \beta} = \left.{ \Delta m^2_{\alpha \beta} L_\rho
\sin^2{2 \theta_{\alpha \beta}}  \over  2 E \cos{2 
\theta_{\alpha \beta}} } 
\right|_{\mbox{res}}~~~.\label{3.9}
\end{equation}
evaluated at the resonance point.
In Eq.(\ref{3.8}) $\theta^m_{\alpha \beta}$ is the effective mixing 
angle in matter \cite{6,7,17}, and is given by
\begin{equation}
\tan{2 \theta^m_{\alpha \beta}} = 
{ 2 \Delta m^2_{\alpha \beta} \sin{2 \theta_{\alpha \beta}} \over 2 \sqrt{2} E 
G_F N_e - \Delta m^2_{\alpha \beta}  \cos{2 \theta_{\alpha \beta}}}~~~.
\label{3.10}
\end{equation}
For NSFP transitions \cite{7,17}
\begin{equation}
P_{\bar{\alpha} \beta} \left(\overline{\nu}_{\alpha} \rightarrow 
\overline{\nu}_{\alpha}\right) =
{1 \over 2} + \left( \frac 12 -
\exp\left\{- { \pi \over 2} \gamma_{\bar{\alpha} \beta} 
F_{\alpha \beta} \right\}
\right) \cos{2 \theta_{\alpha \beta}}
\cos{2 \theta_{\bar{\alpha}\beta}^m}~~~,\label{3.11}
\end{equation}
\begin{equation}
\gamma_{\bar{\alpha}\beta} = \left.{ 8 E \mu_{\alpha \beta}^2 B_{\perp}^2 
L_\rho \over \Delta m^2_{\alpha \beta} }
\right|_{\mbox{res}}~~~.\label{3.12}
\end{equation}
In this case the effective mixing angle in matter is 
\begin{equation}
\tan{2 \theta^m_{\bar{\alpha}\beta}} = { 4 E \mu_{\alpha \beta} 
B_{\perp} \over 2 \sqrt{2} E G_F \left(N_e -N_n\right) -
\Delta m^2_{\alpha \beta}  \cos{2 \theta_{\alpha \beta}}}~~~.
\label{3.13}
\end{equation}
Note that for both the adiabatic parameters $\gamma_{\alpha \beta}$ and 
$\gamma_{\bar{\alpha}\beta}$ of Eq.(\ref{3.9}) and (\ref{3.12}), 
the adiabatic condition \cite{6,7,17} reads
\begin{equation}
\gamma_{\alpha \beta},\gamma_{\bar{\alpha} \beta} >> 1 ~~~.\label{3.14}
\end{equation}
The non adiabatic correction factor $F_{\alpha \beta}$ of 
Eq.s(\ref{3.8}) and (\ref{3.11}),
neglecting non adiabatic effects induced by the magnetic field with
respect to those due to density \cite{7}, is given by \cite{12} 
\begin{eqnarray}
F(\theta_{\alpha \beta}) & \simeq & \left(1 \, - \tan^2(\theta_{\alpha \beta})
\right) \left\{ 1 + \frac{1}{3} \left[ \log\left(1 - 
\tan^2(\theta_{\alpha \beta})\right) \right. \right. 
\nonumber\\
&+& \left. \left. 1 - \frac{1 + \tan^2(\theta_{\alpha \beta})}
{\tan^2(\theta_{\alpha \beta})} \log\left(1+\tan^2(\theta_{\alpha \beta})
\right) \right] \right\}~~~.
\label{3.15bis}
\end{eqnarray}
In terms of survival or transition probabilities, referring to
diagrams (\ref{3.3})--(\ref{3.6}), it is possible to write the
expressions for outcoming neutrino distributions as follows 
\begin{eqnarray}
\nne & = & P\left(\ene\right) \noe + \left[ 1 - P\left(\ene\right)\right]
\nox + P\left(\aene\right) \anoe~~~,\label{3.15}\\
\nnm+\nnt & = & \left[ 1 - P\left(\ene\right)\right] \noe +
\left[ 1 - P\left(\aenae\right)- P\left( \aene\right)\right] \anoe\nonumber\\
&+& \left[P\left(\ene\right) + P\left(\aenae \right)\right] \nox~~~,
\label{3.16}\\
\anne & = & P\left(\aenae \right) \anoe +
\left[ 1 - P\left(\aenae\right)\right] \nox
~~~,\label{3.17}
\end{eqnarray}
where $\nom=\anot=\nox$. Note that to obtain the above expression we
have only used the unitarity and the observation that $P\left(\enae
\right)$ is vanishing (up to the first order in the mixing angle).

According to diagrams (\ref{3.3})--(\ref{3.6}) the survival probabilities
$P\left(\ene\right)$ and $P\left(\aenae\right)$ can be written as
products
of the single survival probabilities at the resonances, namely
\begin{eqnarray}
P\left(\ene\right) & \simeq & P_{e \tau} \left(\ene\right) 
P_{e \mu} \left(\ene\right)~~~,\label{3.18}\\
P\left(\aenae\right) &  \simeq & P_{\bar{e} \tau} \left(\aenae\right) 
P_{\bar{e} \mu} \left(\aenae\right)~~~.\label{3.19}
\end{eqnarray}
while the transition probability $P\left(\aene\right)$ takes the 
expression
\begin{eqnarray}
P\left(\aene\right)  & \simeq & P_{\bar{e} \tau} \left(\aenae\right) 
\left[ 1 - P_{\bar{e} \mu}\left(\aenae\right)\right]
\left[ 1 - P_{e \mu}\left(\ene\right)\right]\nonumber\\
& + & P_{e \mu} \left(\ene\right) 
\left[ 1 - P_{\bar{e} \tau}\left(\aenae\right)\right]
\left[ 1 - P_{e \tau}\left(\ene\right)\right]~~~.
\label{3.20}
\end{eqnarray}
Note that all the expressions (\ref{3.18}--(\ref{3.20}) are obtained
under the assumption that the single resonances of diagrams
(\ref{3.3})--(\ref{3.6}) are well separated. 

\section{Numerical results}

According to the above results (\ref{3.15})--(\ref{3.17}), the
deformed thermal neutrino spectra are obtained once that neutrino
parameters are fixed. 

For the electron--muon sector, the relevant parameters can
be fixed according to the explanation assumed for the solar neutrino
problem. In this paper, since we are interested in a possible scenario
in which the neutrino electromagnetic properties play the essential
role we will choose the NSFP explanation \cite{13}. In this case, the
deficit in the solar neutrino flux is mainly due to the conversion
$\ne \rightarrow \anm$, being assumed Majorana neutrinos, which is the
natural choice occurring in GUT theories where a see-saw mechanism
is at work.\\
The alternative scenario of a pure MSW
explanation was treated by some of the present authors in a
previous paper \cite{10}. 

In the NSFP framework, the values for neutrino parameters able to
reproduce the data are $\Delta m^2_{e \mu}\simeq 10^{-8}~eV^2$ and
$\sin\left(2 \theta_{e \mu} \right) \simeq 0.2$ and $\mu_{e
\mu} \simeq 10^{-11} \mu_B$ \cite{13}. 

Concerning the parameters for the electron--tau sector, they are less
constrained. However, we can fix $\Delta m^2_{e \tau} \simeq
m_{\nu_{\tau}}^2 \simeq 25~eV^2$ in order to be able to identify
$\nu_{\tau}$ has the hot component of dark matter 
\cite{Gelmini}, whereas, for the transition magnetic moment we can 
assume $\mu_{e
\tau}$ to be of the same order of $\mu_{e \mu}$, since, typically, 
the enhancement to the electromagnetic properties is due to physics 
beyond
the electroweak interactions, which hardly distinguishes between
$\tau$--leptons and $\mu$--leptons. Hence, the only remaining
parameters is $\theta_{e \tau}$, for which we will choose three
indicative values, namely, $10^{-1}$, $10^{-4}$ and $10^{-8}$. Values
in this range are for example predicted by SUSY GUT theories \cite{9},
where one could expect in principle an enhancement of the neutrino
electromagnetic properties. 

In Figures 1--3, we show the deformed neutrino distributions for $\nu_e$,
the sum of $\nm$ and $\nt$, and $\ane$ distributions, respectively,
versus their initial
distributions. In all figures, the solid line
represents the initial distribution and the predictions for the
outcoming neutrino distributions are obtained for the above three
values of $\theta_{e \tau}$.  Furthermore, in Figure 4, we plot the
adiabaticity parameters $\gamma_{e \mu}$, $\gamma_{e \tau}$
and $\gamma_{\bar{e} \mu}$, $\gamma_{\bar{e} \tau}$ of eqs.
(\ref{3.9}) and
(\ref{3.12}). In particular the dashed line represents
$\gamma_{e \tau}$ for the two cases $\theta_{e \tau}=10^{-1}$
(upper line) and $\theta_{e \tau}=10^{-4}$ (lower line). 

As one can see from Figure 4, for $\theta_{e \tau}=10^{-1}$ the MSW
$\ne \leftrightarrow \nt$ transition is the only one to be adiabatic,
and thus the $\ne \rightarrow \nt$ conversion is the only one 
which provides an efficient reshuffling of initial distributions.
It corresponds (dashed line in Figure 1) to
a depletion of $\ne$ distribution in favour of $\nt$ (dashed line of
Figure 2). Note, however, that the adiabaticity of MSW transitions
(\ref{3.9}) decreases with neutrino energy, whereas it increase for
NSFP (\ref{3.12}). 

The MSW $\ne \leftrightarrow \nt$ conversion becomes less and less
efficient as $\theta_{e \tau}$ decreases, as can be seen by the other
two lines in Figures 1 and 2. For the other values of $\theta_{e
\tau}$, in fact, $\gamma_{e \tau} << 1$. However, as one can see
from the dashed-dotted line of Figure 1, corresponding to the
extremely small value $\theta_{e \tau}=10^{-8}$, a conversion of
$\nu_{e}$ in other kind of neutrinos still remains. In fact, in this
case the only conversions remaining are the NSFP as one can see by
observing the antineutrinos spectra of Figure 3. Note that since we
have fixed $\mu_{e \mu}\sim\mu_{e \tau}\sim10^{-11}~\mu_B$, all the deformed
antineutrino distributions are almost coincident, because, in this case, 
they are almost independent of the on neutrino mixing angles
(see Eq.s.(\ref{3.11})--(\ref{3.13})). 

The distortions in the neutrino and/or antineutrino energy spectra can
be observed in the future large neutrino detectors, now under
constructions, like SuperKamiokande and SNO. To quantify our
predictions on directly observable quantities, let us consider for
example the $\nu$--$e^-$ elastic scattering for neutrino detection. 
A relevant quantity, which measures the deformation on neutrino
distributions due to resonant transitions is the ratio between the
recoil electron spectrum for deformed neutrino distributions and the
undeformed one. It is defined as
\begin{equation}
R\left(T,E^{th}_{\nu}\right) = {S\left(T,E^{th}_{\nu}\right) \over
S^0\left(T,E^{th}_{\nu}\right) }~~~,
\label{4.1}
\end{equation}
where
\begin{eqnarray}
S\left(T,E^{th}_{\nu}\right) & = & k
\left\{ \int_{E^{th}_{\nu}}^{E_{max}} \left[\nne(E)
{d \sigma \over d T} (\ne e^- \rightarrow \ne e^-) 
+ \anne(E) {d \sigma \over d T} (\ane e^- \rightarrow \ane e^-) \right.\right.
\nonumber\\
& + & \left(\nnm(E)+\nnt(E)\right)
{d \sigma \over d T} (\nm(\nt) e^- \rightarrow \nm(\nt) e^-) 
\nonumber\\
&+& \left. \left.  \left(\annm(E)+\annt(E)\right) 
{d \sigma \over d T} (\anm(\ant) e^- \rightarrow \anm(\ant) e^-) 
\right]~ dE\right\}~~~,
\label{4.2}
\end{eqnarray}
$E^{th}_{\nu}$ being the neutrino energy threshold, while $E_{max}$ is
the endpoint in the neutrino spectra, and
$S^0\left(T,E^{th}_{\nu}\right)$ is defined as (\ref{4.2}), but for
the initial distributions. Note that $R$ is independent of the factor
$k$ which contains the information about the percentage of supernova
neutrino reaching the detector and other reduction factors due to the
detector set up. For the explicit expressions of the differential
cross sections occurring in (\ref{4.2}) see for example
Ref.\cite{okun}. 

In Figure 5, $R$ is shown for the three values of $\theta_{e \tau}$
considered, by assuming for simplicity a vanishing neutrino energy
threshold and $E_{max}=50~MeV$. 

\section{Conclusions}

In this paper we have studied the deformation of thermal neutrino spectra emitted from
a supernova due to resonant transition effects. In
particular we have considered three flavour light Majorana neutrinos,
as suggested by GUT theories like SO(10) for the neutrino degrees of
freedom not taking mass at the intermediate scale ($\sim 10^{11}GeV$),
according to the see-saw mechanism \cite{8}. 

In the present analysis, which represents an extension of a previous
study in which MSW transitions were considered only, we have assumed
enhanced electromagnetic neutrino properties, as suggested by several
unified models and by the NSFP solution to the SNP. Thus, the
supernova outcoming neutrinos experience spin flavour precessions
\cite{nsfp} in addition to the usual MSW phenomena \cite{6}. 

We have calculated the deformed neutrino spectra once neutrino
parameters, like masses, mixing angles and transitional magnetic
moments are fixed (CPT invariance forbids diagonal magnetic moments
for Majorana neutrinos). In this scenario, since we are interested in
a situation in which the enhanced electromagnetic moments play an
essential role in neutrino physics, we assume the results obtained in
Ref.\cite{13}, where the deficit in the solar neutrino flux is
explained in terms of the NSFP conversion $\ne \rightarrow \anm$.
Thus, from the experimental data one can obtain all the parameters of
$e$--$\mu$ sector. As far as the $e$--$\tau$ sector is concerned, the
magnetic transitional moment $\mu_{e \tau}$  can be reasonably fixed
to be of the order of $\mu_{e \mu}$, since the enhancement should be
due to physics beyond the electroweak interaction and thus 
it is reasonable to assume that it should contribute in a comparable way to all leptonic families.
Furthermore,
requiring that $\nt$ is the hot component of DM \cite{Gelmini}, its
mass can be fixed equal to $5~eV$. 

For three choices of $\ne$--$\nt$ mixing angle, we have then
explicitly obtained the deformed $\ne$, $\nm+\nt$ and $\ane$ energy
spectra, as reported in Figures 1--3. Finally, our predictions for a
directly observable quantity, such as the ratio $R$ defined in
Eq.(\ref{4.1}) have been given in Figure 5. 

\noindent
{\bf \Large Acknowledgements}\\
We thank Prof. F. Buccella for valuable comments and remarks.

\newpage

\begin{figure}[]
\epsfysize=18cm
\epsfxsize=16cm
\epsffile{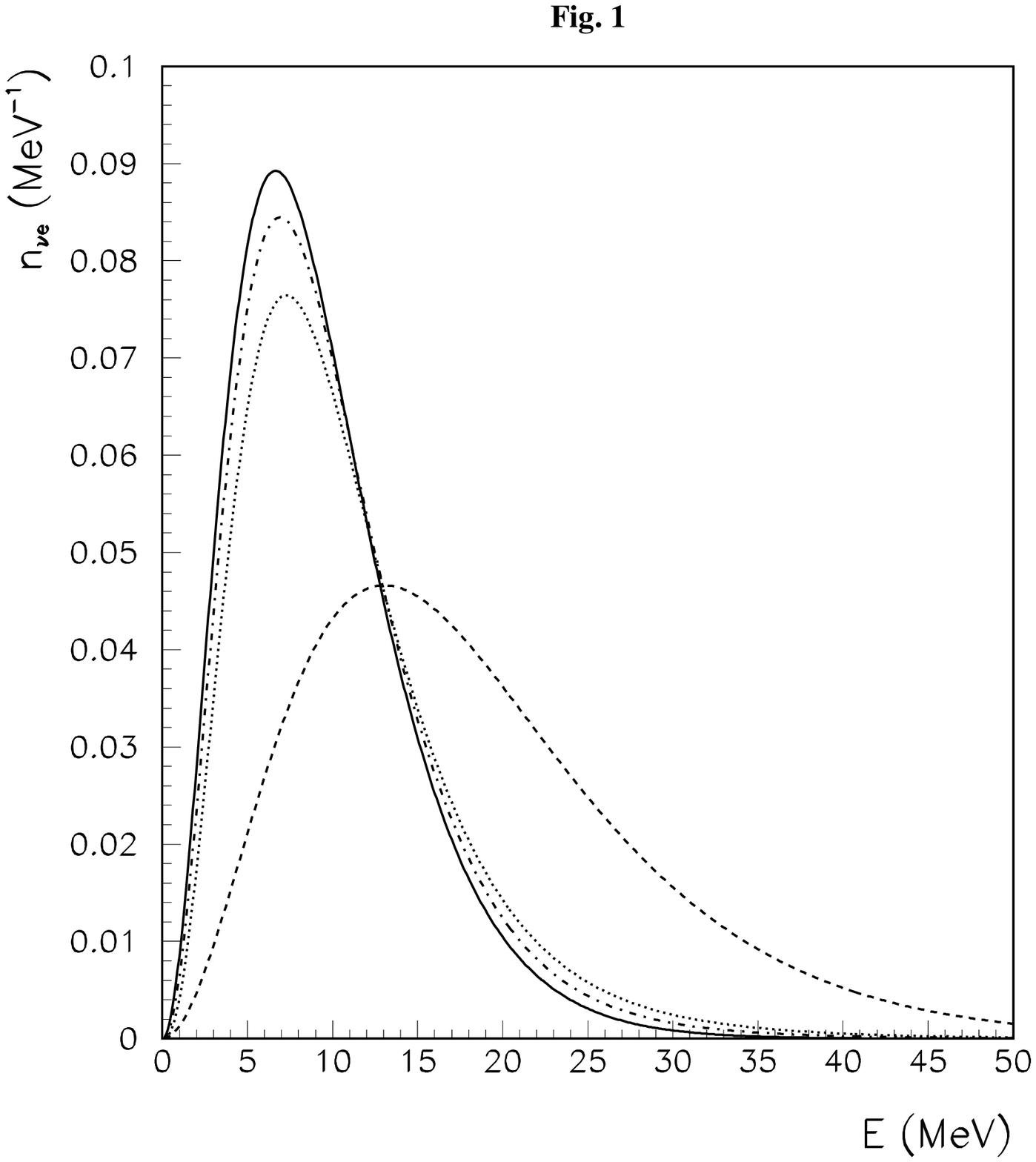}
\caption[]{The energy spectra for $\ne$ versus energy are reported.  
The solid line
represents the initial distribution, whereas the dashed line corresponds
to the distorted spectra for $\theta_{e \tau}=10^{-1}$. The dotted line
and dashed--dotted correspond to $\theta_{e \tau}=10^{-4}$ and 
$10^{-8}$, respectively.}
\end{figure} 

\newpage

\begin{figure}[]
\epsfysize=18cm
\epsfxsize=16cm
\epsffile{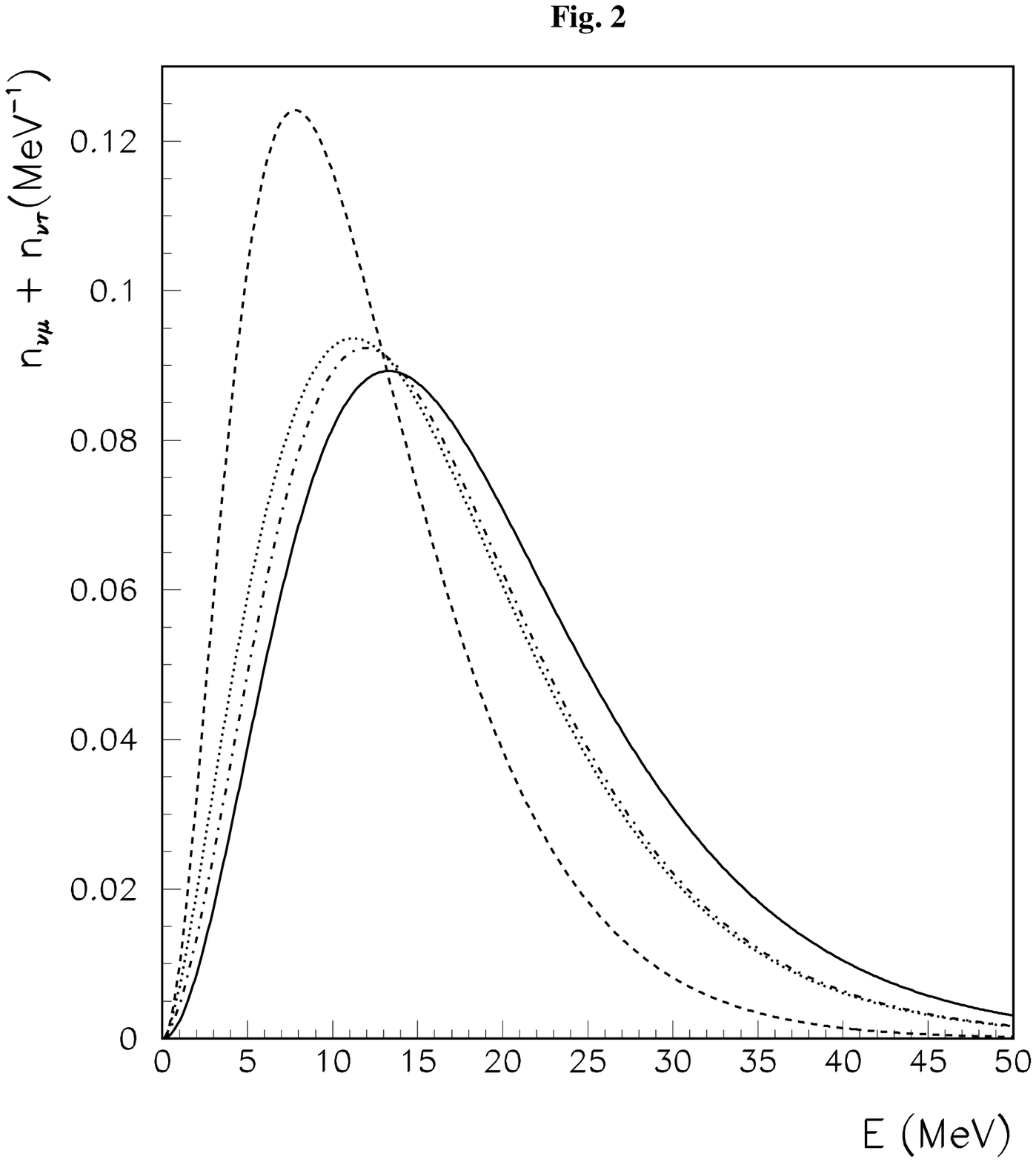}
\caption[]{The energy spectra for $\nnm + \nnt$,
with the same notation of Figure 1,}
\end{figure} 

\newpage

\begin{figure}[]
\epsfysize=18cm
\epsfxsize=16cm
\epsffile{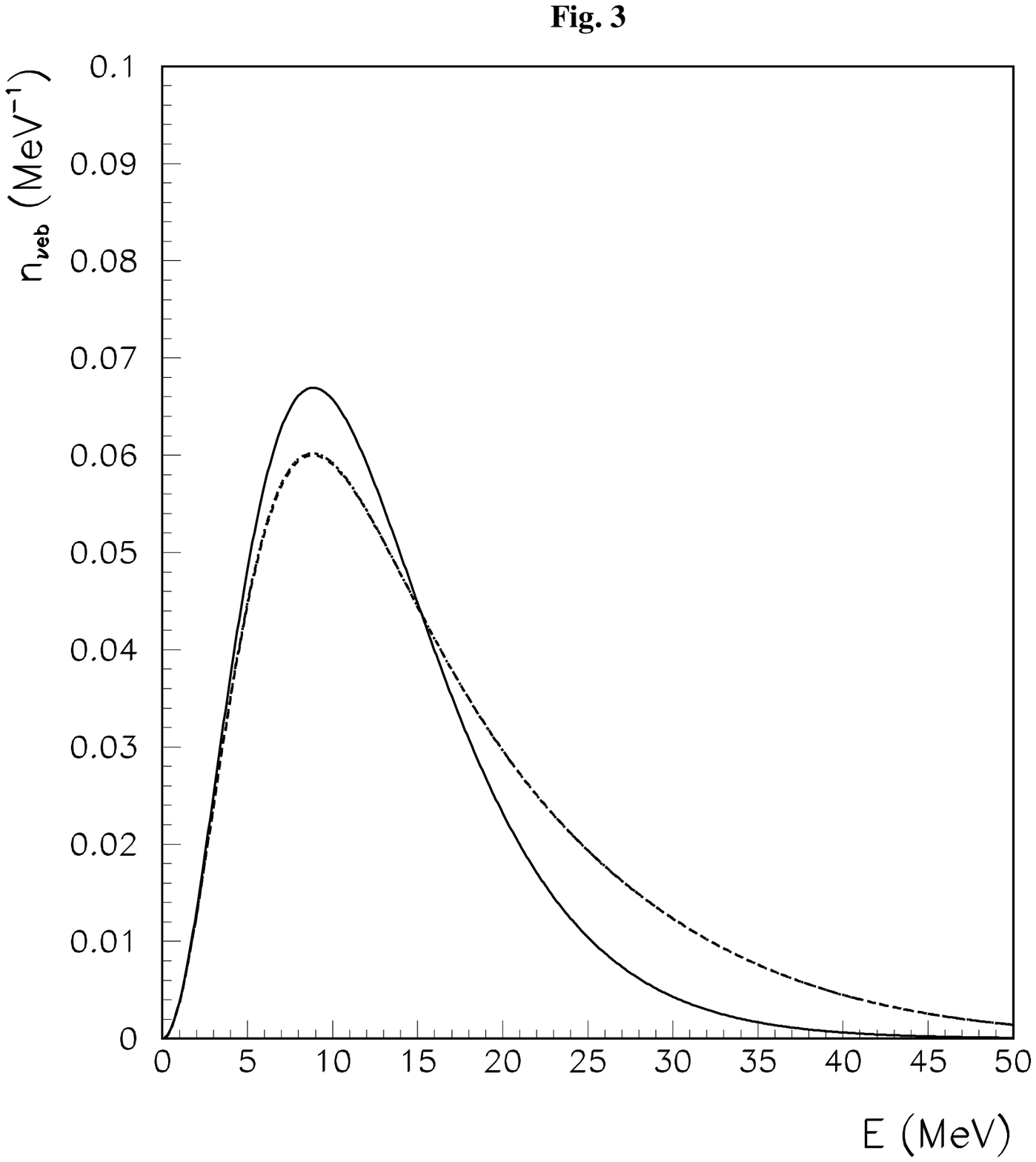}
\caption[]{The energy spectra for
$\anne $, with the same notation of Figure 1}
\end{figure} 

\newpage

\begin{figure}[]
\epsfysize=18cm
\epsfxsize=16cm
\epsffile{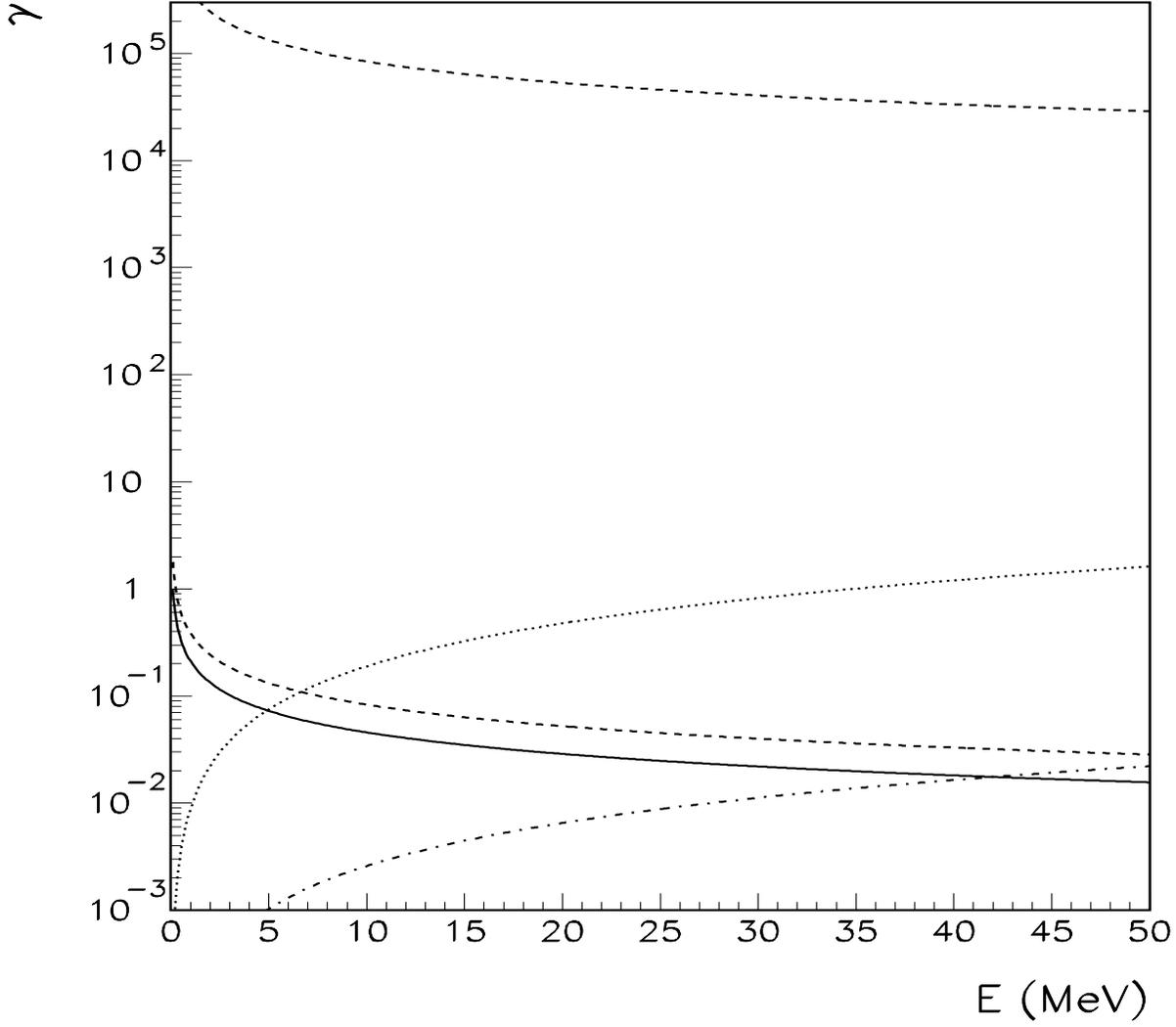}
\caption[]{Energy dependence of the adiabaticity parameters 
$\gamma_{\alpha
\beta}$  and $\gamma_{\bar{\alpha} \beta}$ 
of eqs. (\ref{3.9}) and (\ref{3.12}).
The solid line represents $\gamma_{e \mu}$ and the dashed
lines $\gamma_{e \tau}$: the upper one corresponds to $\theta_{e
\tau}=10^{-1}$ and the lower to $\theta_{e \tau}=10^{-4}$. The dotted
line is $\gamma_{\bar{e} \mu}$ and the dashed--dotted represents
$\gamma_{\bar{e} \tau}$.} 
\end{figure} 

\newpage

\begin{figure}[]
\epsfysize=18cm
\epsfxsize=16cm
\epsffile{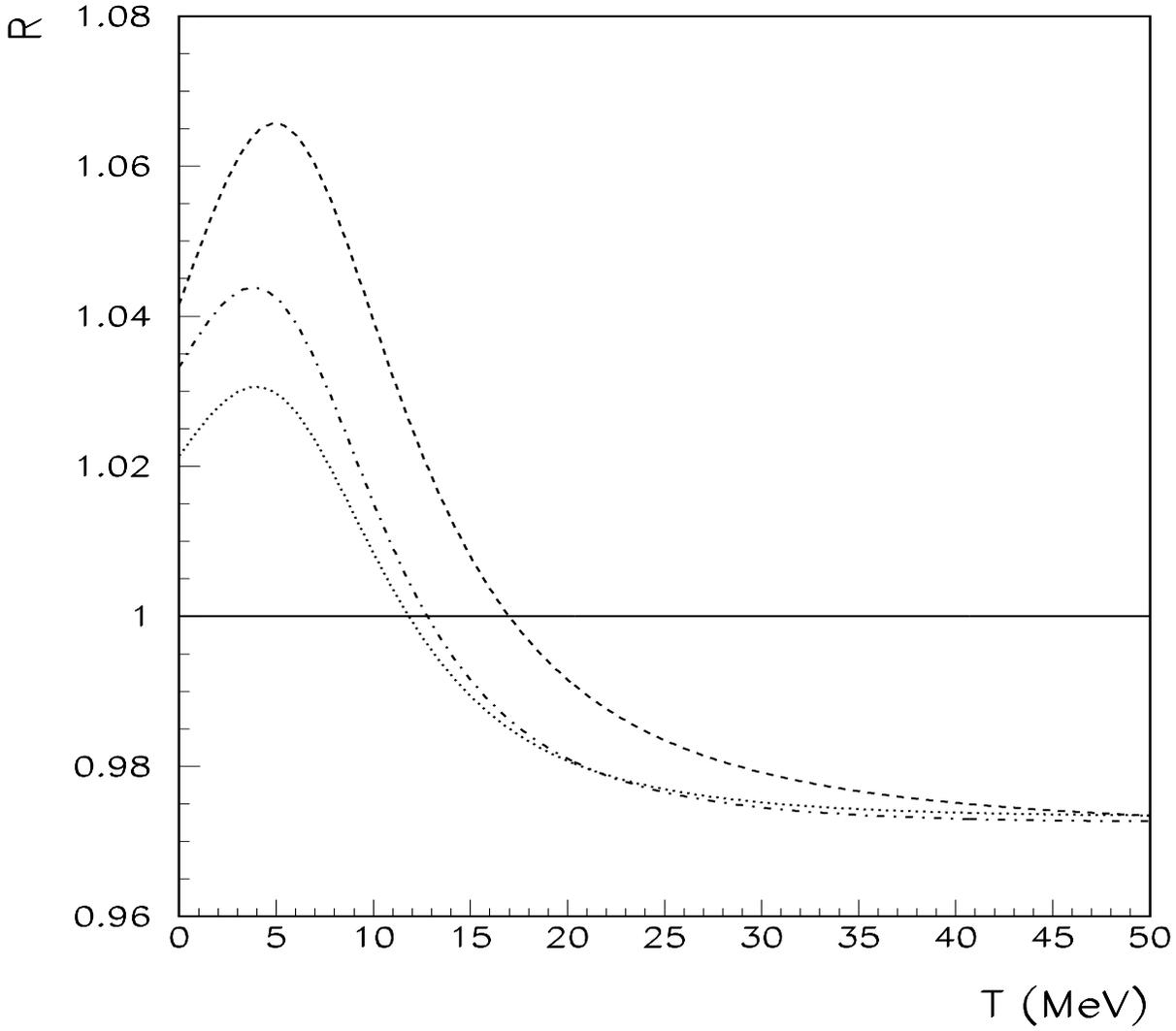}
\caption[]{The ratio R (see Eq.(\ref{4.1})) is plotted for $\theta_{e
\tau}=10^{-1}$ (dashed line), $10^{-4}$ (dotted line) and $10^{-8}$
(dashed--dotted line). For simplicity a vanishing neutrino energy
threshold has been used.} 
\end{figure} 

\end{document}